# Nearest Neighbor Value Interpolation

Olivier Rukundo
Department of Electronics and Information Engineering
Huazhong University of Science and Technology, HUST
Wuhan, China

Hanqiang Cao
Department of Electronics and Information Engineering
Huazhong University of Science and Technology, HUST
Wuhan, China

*Abstract*—**This paper presents the nearest neighbor value (NNV) algorithm for high resolution (H.R.) image interpolation. The difference between the proposed algorithm and conventional nearest neighbor algorithm is that the concept applied, to estimate the missing pixel value, is guided by the nearest value rather than the distance. In other words, the proposed concept selects one pixel, among four directly surrounding the empty location, whose value is almost equal to the value generated by the conventional bilinear interpolation algorithm. The proposed method demonstrated higher performances in terms of H.R. when compared to the conventional interpolation algorithms mentioned.**

## I. Introduction

Image interpolation is the process by which a small image is made larger by increasing the number of pixels comprising the small image [1]. This process has been a problem of prime importance in many fields due to its wide application in satellite imagery, biomedical imaging, and particularly in military and consumer electronics domains. At an early stage of research, non-adaptive methods such as nearest, bilinear and bicubic interpolation methods were developed for digital image interpolation purposes. Those traditional methods were markedly different in image resolution, speed, and theoretical assumptions (i.e. theory of spatial variability) [2], [3]. To the best of my knowledge, most of the assumptions applied today reduce interpolated image resolution due to the lowpass filtering process involved into their new value creative operations [4]. However, the nearest neighbor assumption does not permit to create a new value, instead set the value at the empty location by replicating the pixel value located at the shortest distance. The effect of this is to make image pixel bigger which results in heavy jagged edges thus making this algorithm more inappropriate for applications requiring a H.R image (to accomplish certain tasks). A solution to such jaggedness was achieved through the bilinear interpolation [5]. A bilinear based algorithm generates softer images but blurred thus making the algorithm inappropriate also for H.R. applications. The blurredness problem was reduced by introducing the convolution based techniques [6]. Such algorithms performed better than the two in terms of the visual quality but are also inappropriate to use where the speed is of the prime importance. Now, since the source image resolution is often reduced after undergoing the interpolation process, the easy way to generate a H.R. image using linear interpolation means is to reduce, at any cost, any operation that would underestimate or overestimate some parts of the image. In other words, it would be better if we could avoid 100% any operation leading to the new pixel value creation for image interpolation purposes. In this regards, one way to reduce using the newly created values, is based on supposing that one, of the four pixels, has a value that is appropriate enough to be assigned directly at an empty location. The problem, here, is to know which one is more appropriate than the others or their weighted average, etc. Therefore, we propose a scheme to be guided, throughout our H.R. interpolated image search, by the value generated by the conventional bilinear interpolator. The Fig.2 briefly explains how this can be achieved. More details are given in Part III.

This paper is organized as follows. Part II gives the background, Part III presents the proposed method, Part IV presents the experimental results and discussions and Part V gives the conclusions and recommendation.

## II. Background

The rule in image interpolation is to use a source image as the reference to construct a new or interpolated/scaled image. The size of the new or constructed image depends on the interpolation ratio selected or set. When performing a digital image interpolation, we are actually creating empty spaces in the source image and filling in them with the appropriate pixel values [2]. This makes the interpolation algorithms yielding different results depending on the concept used to guess those values. For example, in the nearest neighbor technique, the empty spaces will be filled in with the nearest neighboring pixel value, hence the name [3]. This (nearest neighbor algorithm) concept is very useful when speed is the main concern. Unlike simple nearest neighbor, other techniques use interpolation of neighboring pixels (while others use the convolution or adaptive interpolation concepts - but these two are beyond the scope of this paper), resulting in smoother image. A good example of a computationally efficient basic resampling concept or technique is the bilinear interpolation. Here, the key idea is to perform linear interpolation first in one direction, and then again in the other direction. Although each step is linear in the sampled values and in the position, the interpolation as a whole is not linear but rather quadratic in the sample location [5]. In other words, the bilinear interpolation algorithm creates a weighted average value that uses to fill in the empty spaces. This provides better tradeoff between image quality and computational cost but blurs the interpolated image thus reducing its resolution.

## III. PROPOSED METHOD

Assume that the letters A, K, P and G represent the four neighbors and E represents the empty location value as shown in Fig.1.

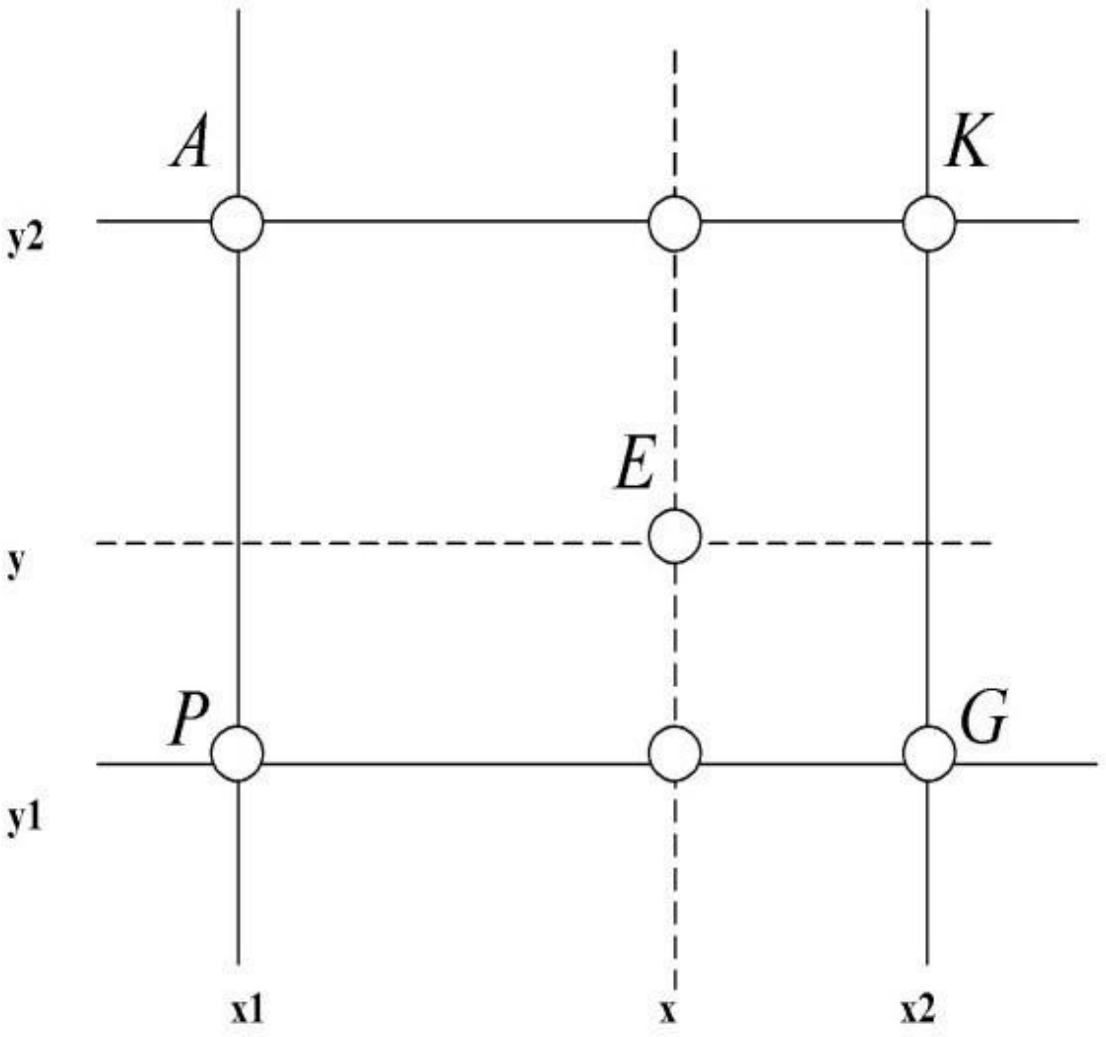

Fig. 1: Four neighbors locations around an empty location E

In order to implement successfully the proposed scheme, the following steps have been respected.

### A. Neighbors mode calculation

Let us call $L \;\; [A, K, P, G]$ a set of four neighbors or data. In statistics, the mode is the value that occurs most frequently in a data set or a probability distribution [7]. Here, the first step is to check whether in $L$ there is a mode or not (i.e. if there exists a mode in $L$ ). If the mode exists then, the empty location will be assigned that mode. If the mode does not exist in $L$ (i.e., when two data in $L$ appear the same number of times or when none of the $L$ data repeat) then, we proceed to performing the bilinear interpolation among $L$ data in order to achieve a bilinearly interpolated value or bilinear value. Once the bilinear value is obtained, we do the subtraction operations as shown by Eq.(1), Eq.(2), Eq.(3) and Eq.(4). The letter B represents the bilinear value.

$$\left| A \;\; B \right| \;\; V_1 \tag{1}$$

$$\left| K \;\; B \right| \;\; V_2 \tag{2}$$

$$\left| P \;\; B \right| \;\; V_3 \tag{3}$$

$$\left| G \;\; B \right| \;\; V_4 \tag{4}$$

The values obtained, from the subtraction operations, are absolute values and can be represented by $V_1, V_2, V_3$ and $V_4$. Before, we proceed to finding the pixel value yielding the minimum difference, we must check that none of these absolute values is equal to another or simply occurs most frequently. In other words, we must check again the mode so that we can be able to end up with one neighbor whose value is nearly equal to the value yielded by the bilinear interpolator.

### B. Absolute differences mode calculation

At this stage, the mode is calculated from a given set $J$ containing all the absolute differences $J \;\; [V_1, V_2, V_3, V_4]$. If there exists a mode in $J$ then, we can find out that the mode is the minimum value or not, before we can proceed further. For instance, consider the following three examples.

*Example 1*: $V_1 \;\; 0.2, V_2 \;\; 0.2, V_3 \;\; 0.2, V_4 \;\; 0.8$

In this example, the mode is 0.2 and 0.2 is the minimum value. So, in order to avoid the confusion our algorithm will only consider/select the first value from $J$. The selection of the first value can be achieved based on the subscripted indexing theory [8]. Once this value is selected, we can calculate the absolute difference between this value and bilinear value and the difference obtained is assigned to the empty location.

*Example 2*: $V_1 \;\; 0.2, V_2 \;\; 0.8, V_3 \;\; 0.8, V_4 \;\; 0.8$

In this example, the mode is 0.8 and unfortunately 0.8 is not the minimum value therefore our concept, which is directed by the minimum difference value between the value yielded by the bilinear and one of the four neighbors value, cannot be respected. To solve this issue, first of all we find the value that is less or equal to the mode. In the matlab the *find* function returns indices and values of nonzero elements [9]. The obtained elements are presented in ascending order. In this example, the value that is less or equal to the mode would be 0.8 or 0.2 but since the two values cannot be selected at the same time, we can pick the first minimum value by applying again the subscripted index method. Once the minimum value is obtained, we can find the neighbor that corresponds to that minimum value and calculate the absolute difference between that value and bilinear value, then assign it to the empty location.

### C. When there is no 'absolute differences' mode

This case can also be regarded as an example number three of the B part (even though it is presented in C part).

*Example 3*: $V_1 \;\; 0.2, V_2 \;\; 0.2, V_3 \;\; 0.8, V_4 \;\; 0.8$

As shown, in this example, there is no mode when two data/elements of a set repeat the same number of times. The same when all $J$ elements are different. In both cases, we have to find the minimum value in $J$ using matlab *min* function so that we can be able to apply the subscripted indexing to get the first minimum value. Once the minimum value is obtained, we must find the neighbor that corresponds to that minimum number. This can be achieved by subtracting the minimum difference from the bilinear value, because the minimum value is equal to the neighbor value minus the bilinear value (see Eq.(1), Eq.(2), Eq.(3) and Eq.(4) ). Then, the obtained value is assigned to the empty location.

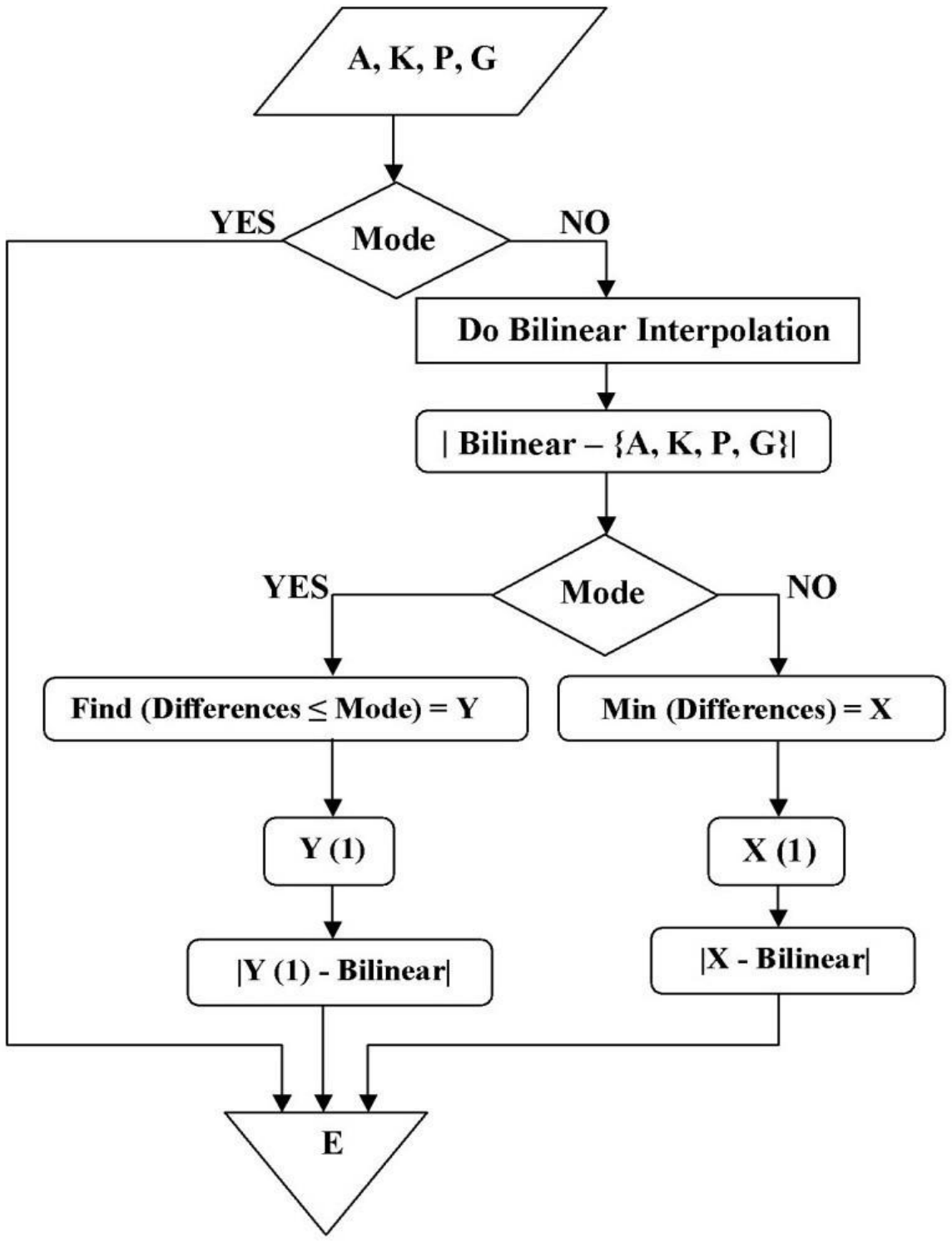

Fig.2: The summary of the proposed method

The Fig.2 shows four data input. These data are in fact the four neighbors surrounding the empty location. The $E$ destination represents the final interpolated value that must be assigned to the empty location, accordingly. The aim, of finding this value in this way, is to minimize the underestimation or overestimation of some parts of image texture after undergoing the interpolation process because of the problems caused by the lowpass filtering processes involved in many linear interpolators, bilinear in particular.

## IV. EXPERIMENTS AND DISCUSSIONS

We tested the proposed NNV algorithm for image details quality (i.e. H.R), Matlab-lines Execution Time (MET) and Peak Signal to Noise Ratio (PSNR) against the conventional nearest, bilinear and bicubic interpolation algorithms using four full grayscale images shown in Fig.3. The interpolated images (ratio n = 4 and n = 2) are shown in Fig.[4-11]. The corresponding MET and PSNR results are shown in the Table 1 and Table 2. A higher peak signal to noise ratio would normally indicate the higher quality of the output image. The PSNR can easily be defined via the Mean Squared Error where one of the monochrome images $I$ and $K$ is considered as a noisy approximation of the other.

$$MSE = \frac{1}{mn}\sum_{i=0}^{m-1}\sum_{j=0}^{n-1}\left[I(i,j) - K(i,j)\right]^2 \qquad (5)$$

The PSNR is defined as:

$$PSNR = 10 \log_{10}\left(\frac{MAX_I^2}{MSE}\right) = 20 \log_{10}\left(\frac{MAX_I}{\sqrt{MSE}}\right) \qquad (6)$$

where, $MAX_I$ represents the maximum image pixel value. Typically, the PSNR values in lossy image and video compression range from 30 to 50 dB. When the two images are identical, the MSE=0 and consequently the PSNR is undefined.

### A. Original full images - 128 x 128

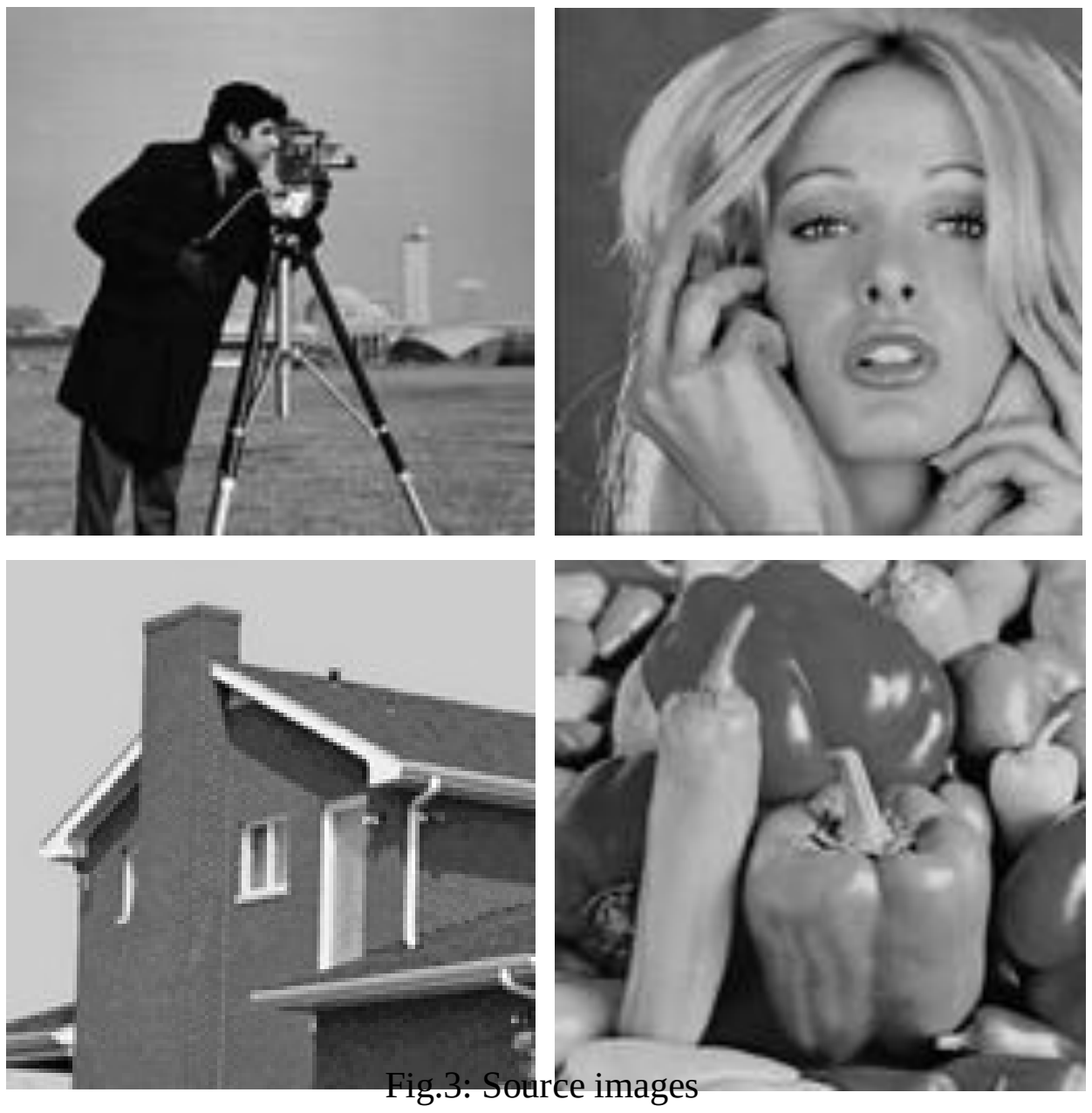

Fig.3: Source images

From left to right - top to bottom - there are Cameraman, Girl, House and Peppers grayscale images. Each image has the size of 128 x 128 and will be interpolated at the ratio n=4 and n=2 in part B and C, respectively.

### B. Full image interpolation & Ratio = 4

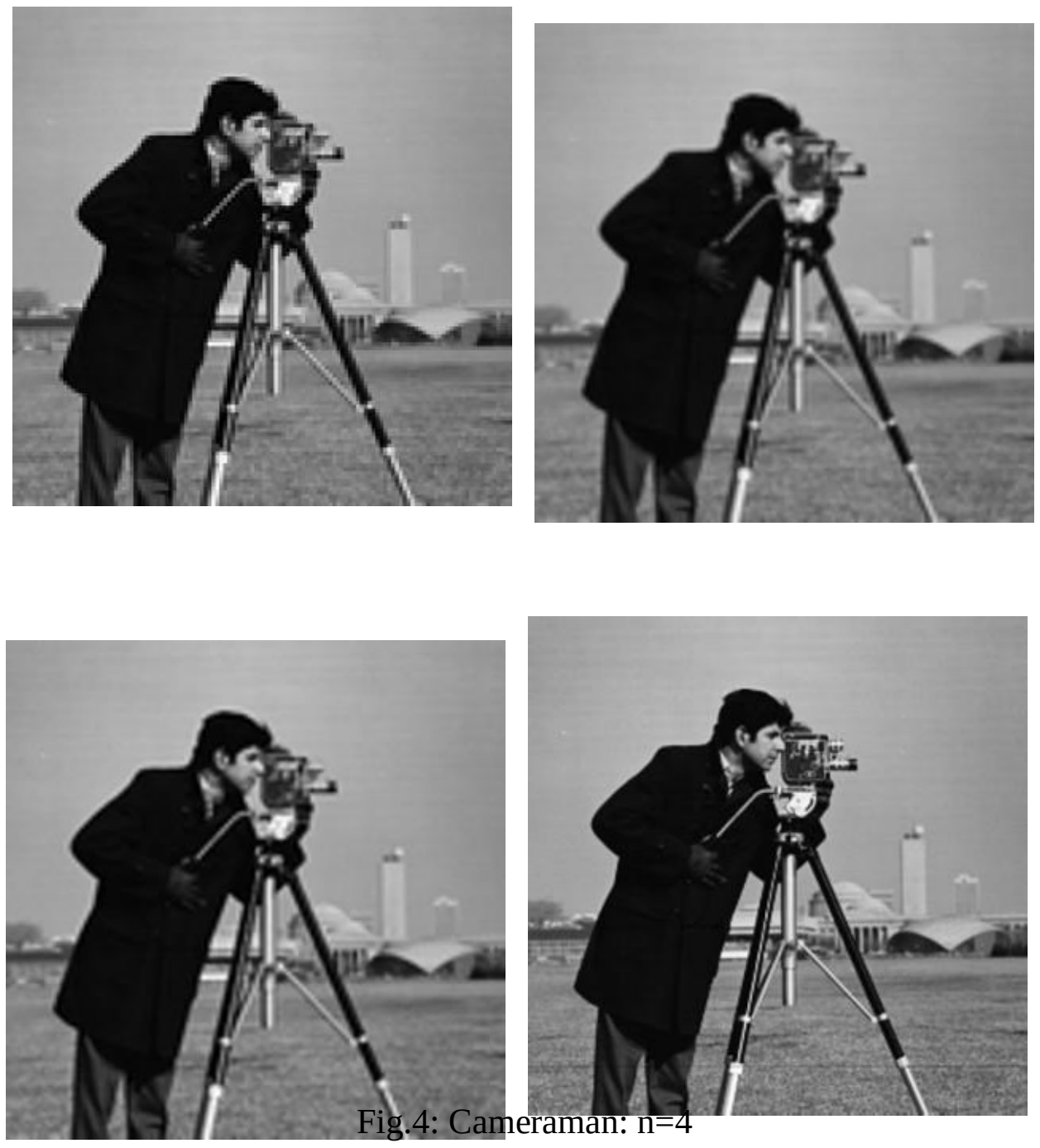

Fig.4: Cameraman: n=4

From left to right - top to bottom – 512 x 512 Cameraman- image interpolated by nearest neighbor (NN), bilinear (Bil.), bicubic (Bic.) and nearest neighbor value (NNV) algorithms, respectively.

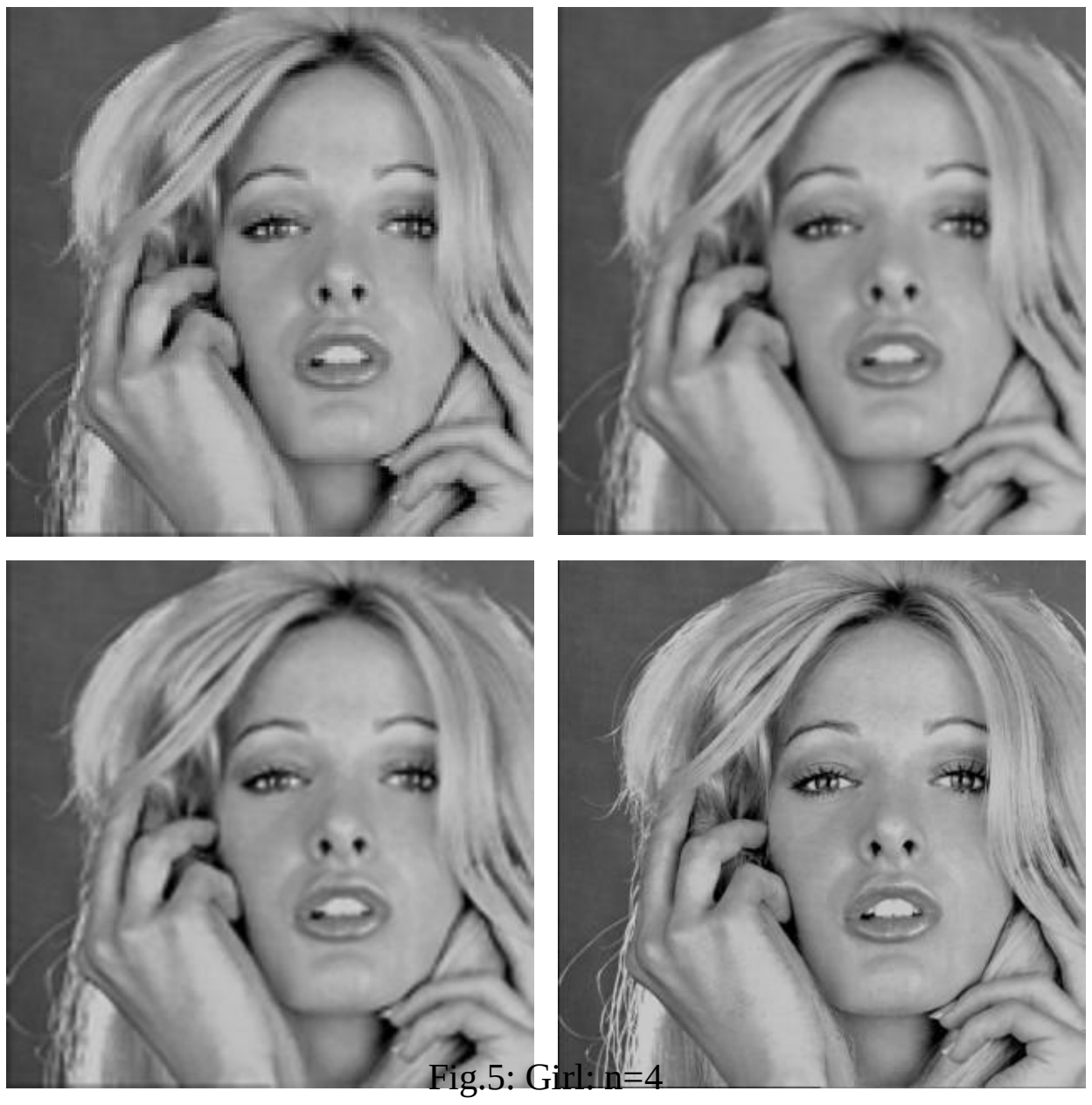

Fig.5: Girl: n=4

From left to right - top to bottom – 512 x 512 Girl-image interpolated by NN, Bil., Bic. and NNV algorithms, respectively.

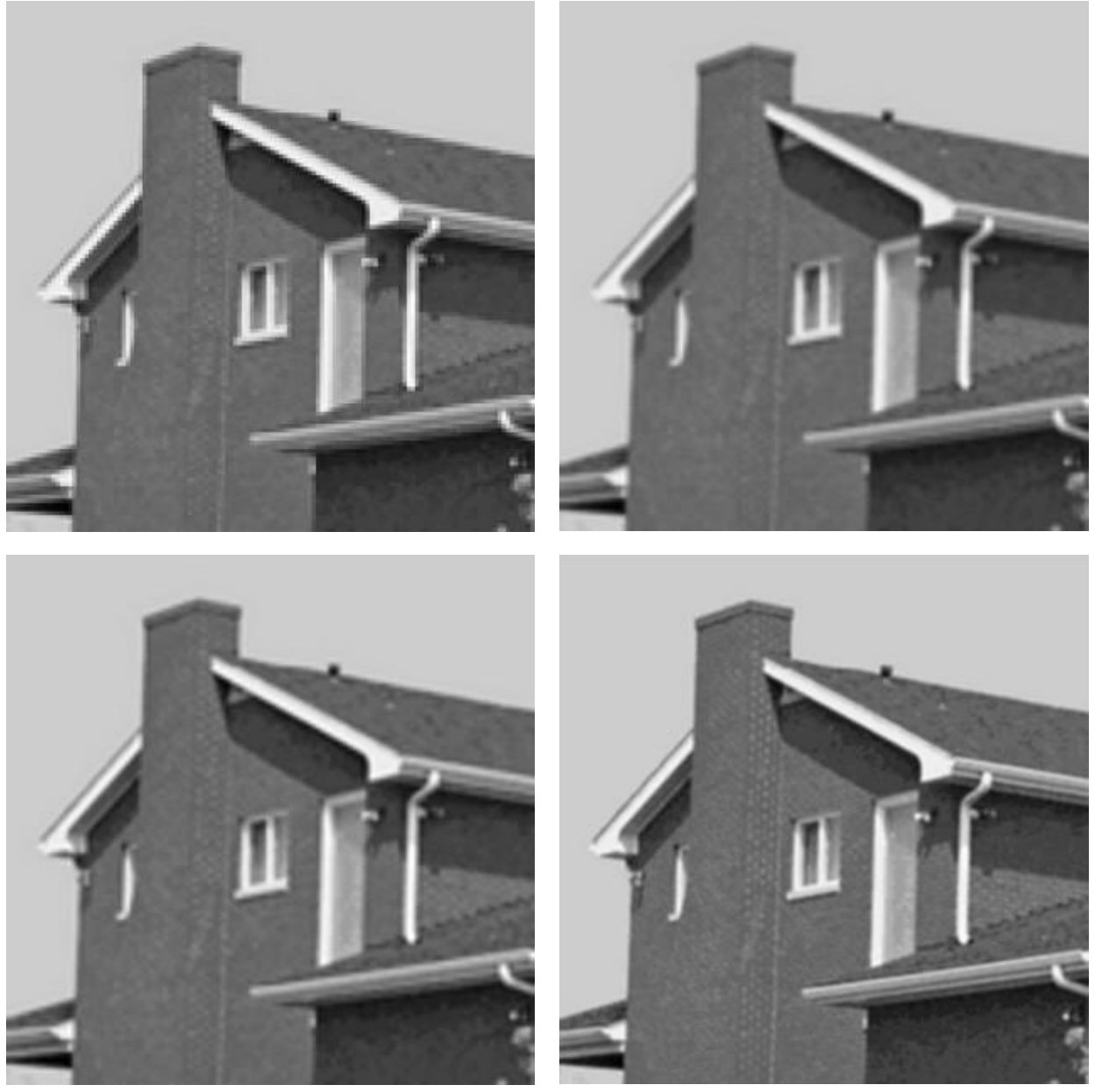

Fig.6: House: n=4

From left to right - top to bottom – 512 x 512 House-image interpolated by NN, Bil., Bic. and NNV algorithms, respectively.

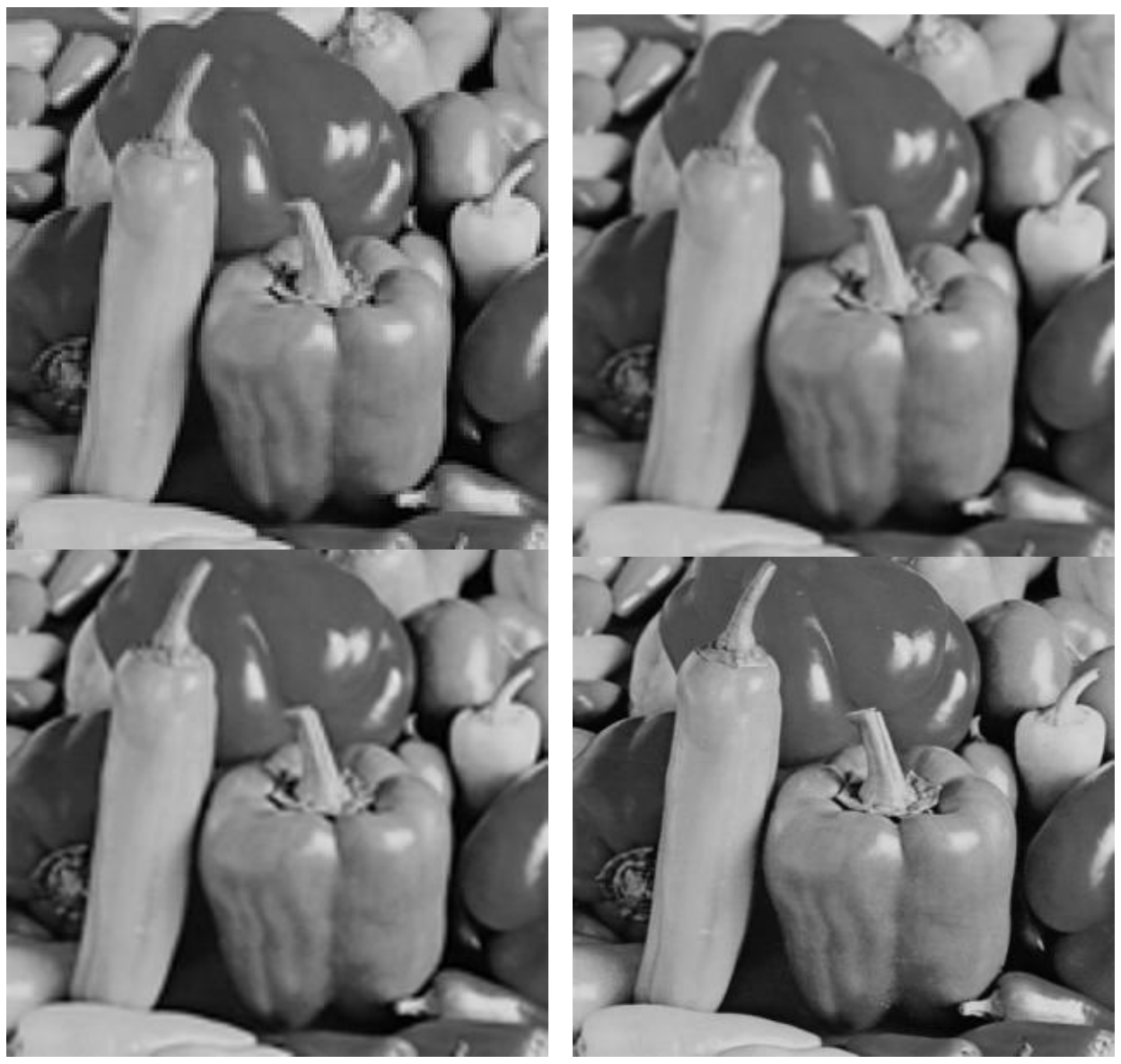

Fig.7: Peppers: n=4

From left to right - top to bottom – 512 x 512 Peppers-image interpolated by NN, Bil., Bic. and NNV algorithms, respectively.

### *C. Full image interpolation & Ratio = 2*

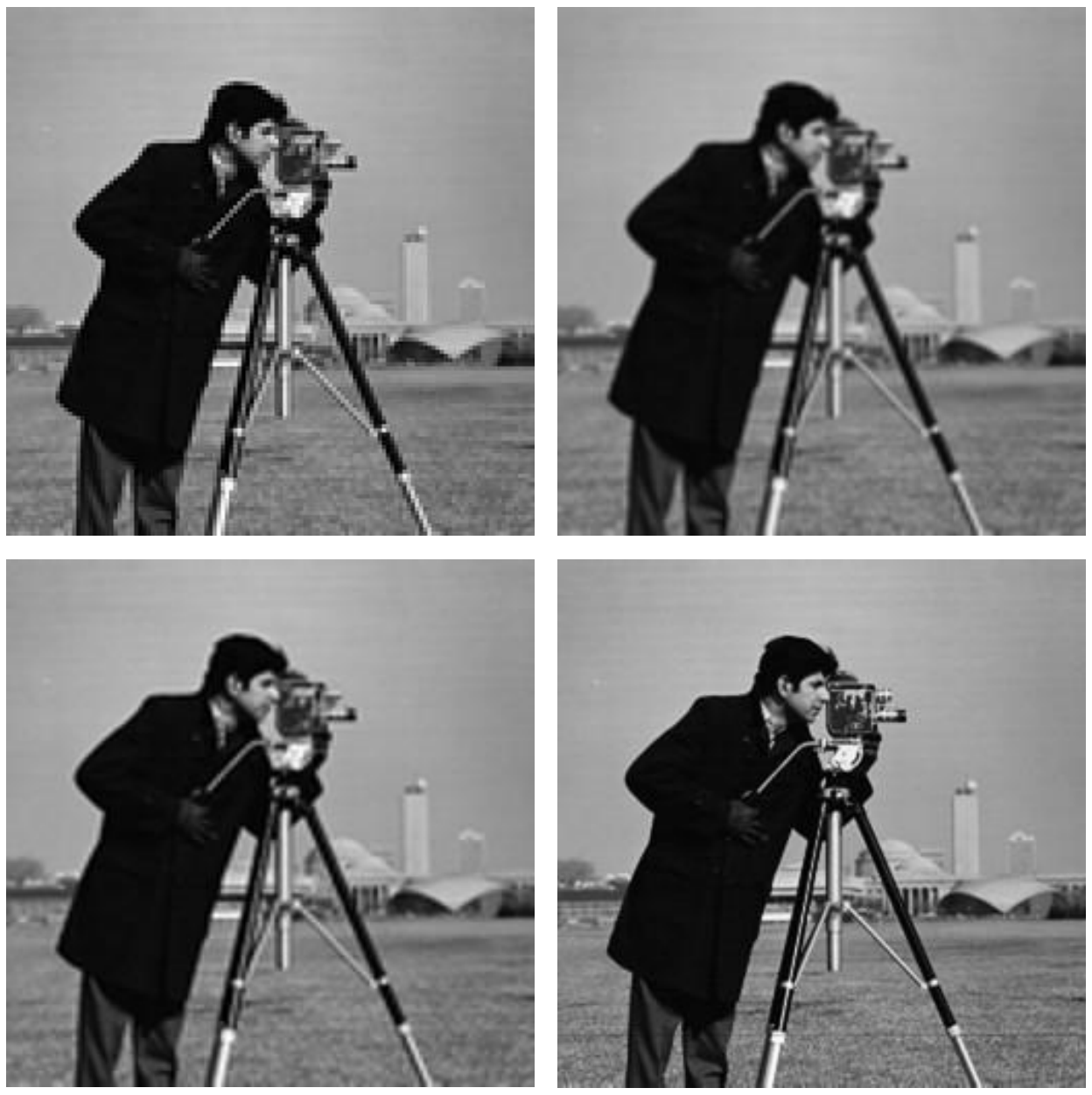

Fig.8: Cameraman: n=2

From left to right - top to bottom – 256 x 256 Cameraman- image interpolated by NN, Bil., Bic. and NNV algorithms, respectively.

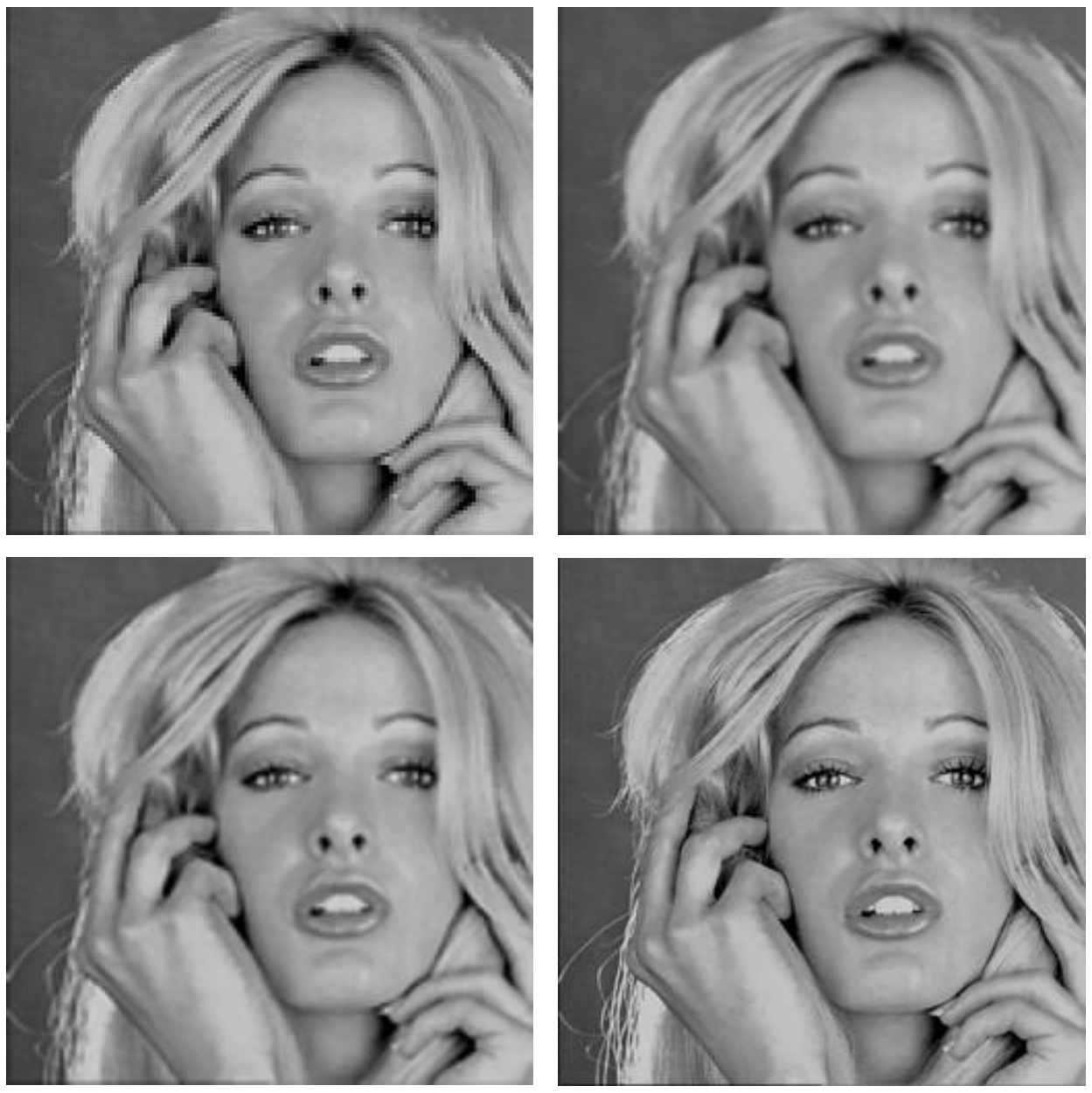

Fig.9: Girl: n=2

From left to right - top to bottom – 256 x 256 Girl-image interpolated by NN, Bil., Bic. and NNV algorithms, respectively.

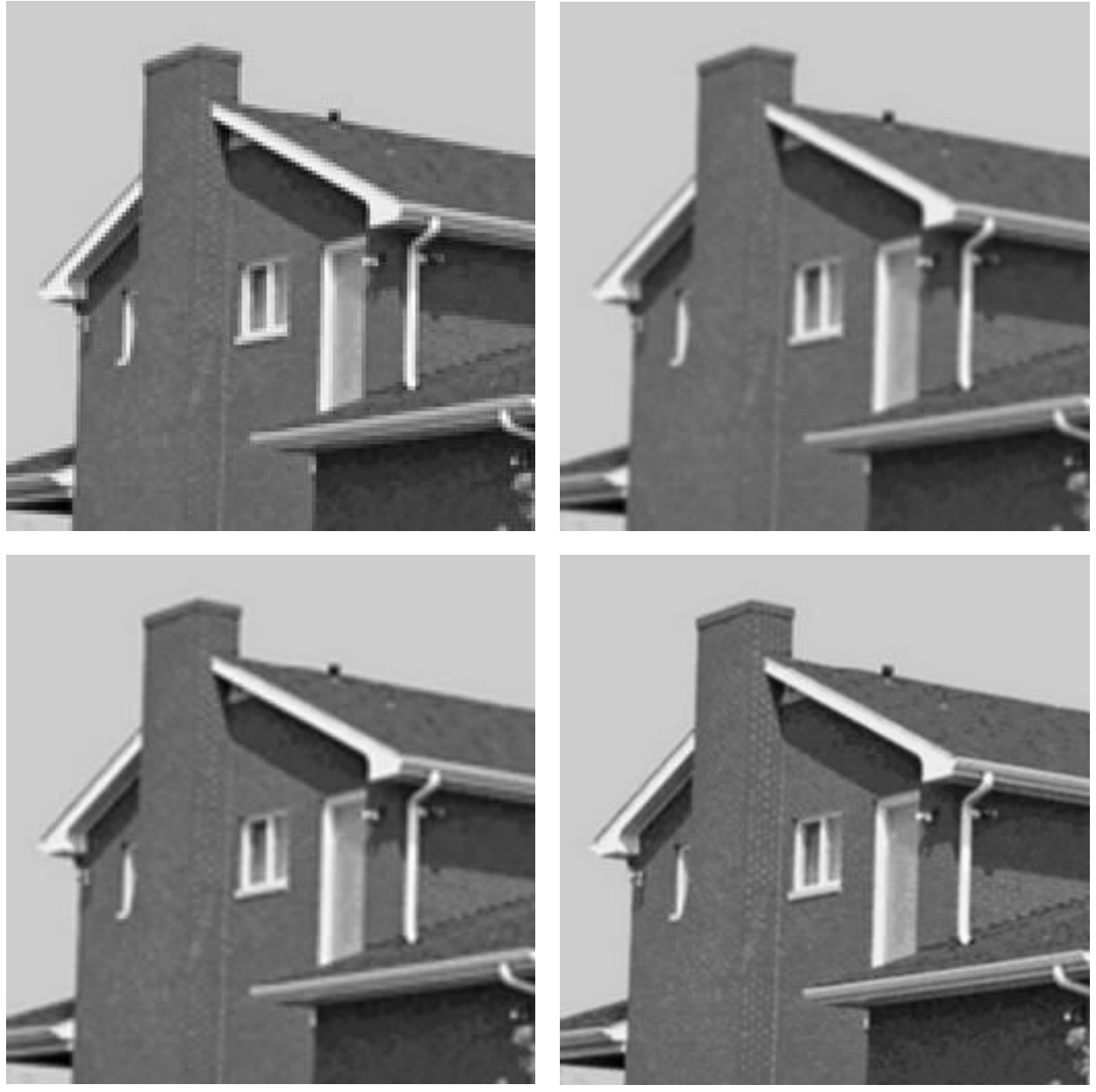

Fig.10: House: n=2

From left to right - top to bottom – 256 x 256 House-image interpolated by NN, Bil., Bic. and NNV algorithms, respectively.

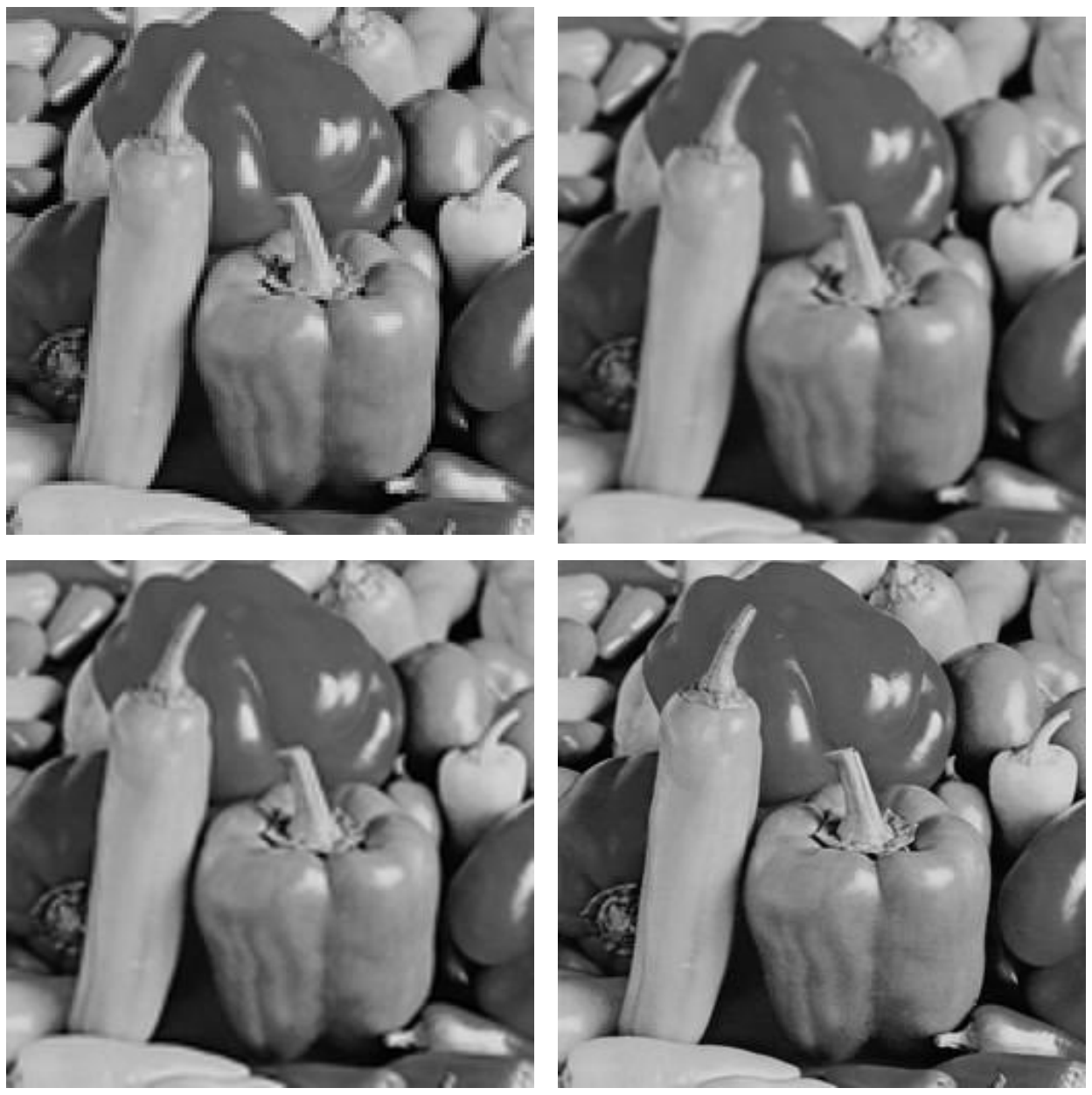

Fig.11: Peppers: n=2

From left to right - top to bottom – 256 x 256 Peppers-image interpolated by NN, Bil., Bic. and NNV algorithms, respectively.

## V. CONCLUSION AND RECOMMENDATIONS

Image interpolation based on the nearest neighbor value has been presented in this paper. The details on how it was developed (i.e. the scheme used) have been presented in part III and the working procedure has been summarized and shown in Fig.2. The MET and PSNR results have been presented in Table 1 and Table 2. Depending on the interpolation ratio selected or set (i.e. depending on the final size desired/targeted), the interpolation algorithms, mentioned here, gave different MET and PSNR as well as visual quality. For example, let us observe the interpolated images shown in part B (i.e. image that were interpolated at the ratio = 4). Starting from, the Cameraman image on the first row, the first image shows a texture with edge jaggedness (i.e. image interpolated using the NNI algorithm) while the second one (i.e. image interpolated using the bilinear algorithm) shows soft but blurred texture. The first image tends to look sharper than the second one. That look difference was noticed due to the lowpass filtering process involved in the algorithm used to interpolate the latter. On the second row, the first image (i.e. image interpolated using the bicubic algorithm) shows smoother but sharper texture so is the second one (i.e. image interpolated using the NNV algorithm), except that the latter shows more readily the image details. The same conclusions can be drawn for other image cases but with a slight change because the best interpolation method for an image may depend on the image itself. In other words, one shoe may not fit all. For the interpolated images shown in part C (i.e. image that were interpolated at the ratio = 2). Except the first image on the first row (i.e. image interpolated using the NNI algorithm), it is difficult to notice the visual differences because the differences were minor (with the exception of the NNV) and it is often problematic as to which one looks the best.

Table 1: PSNR and MET after interpolation & (ratio = 4) 1: Cameraman, 2: Girl, 3: House, 4: Peppers

| | PSNR (dB) | | | | MET (s) | | | |
|---|---|---|---|---|---|---|---|---|
| | NN | Bil. | Bic. | NNV | NN | Bil. | Bic. | NNV |
| 1 | 34.0829 | 34.1135 | 34.1628 | 35.0154 | 0.036866 | 0.058984 | 0.060625 | 0.843483 |
| 2 | 32.9235 | 33.1043 | 33.1655 | 34.2262 | 0.037365 | 0.059842 | 0.060477 | 0.818222 |
| 3 | 35.4771 | 35.4563 | 35.4890 | 36.2054 | 0.038078 | 0.057881 | 0.074557 | 0.788410 |
| 4 | 33.6570 | 33.7862 | 33.8349 | 34.5064 | 0.040556 | 0.062336 | 0.063787 | 0.800693 |

Therefore, the PSNR was used to prove through the calculations which one has really higher quality than the others. As shown in Table 1 and Table 2, the PSNR value generated by the NNV has always been superior to other image interpolators'. This can be interpreted as an overwhelming sign showing the higher performances of the proposed algorithm with respect to other interpolation algorithms mentioned.

Table 2: PSNR and MET after interpolation & (ratio = 2) 1: Cameraman, 2: Girl, 3: House, 4: Peppers

| | PSNR (dB) | | | | MET (s) | | | |
|---|---|---|---|---|---|---|---|---|
| | NN | Bil. | Bic. | NNV | NN | Bil. | Bic. | NNV |
| 1 | 34.2996 | 34.0658 | 34.3385 | 36.0238 | 0.042512 | 0.053586 | 0.055415 | 0.700002 |
| 2 | 33.7806 | 33.8934 | 34.2070 | 35.9882 | 0.037528 | 0.060694 | 0.060948 | 0.780222 |
| 3 | 35.9944 | 35.6859 | 35.9841 | 36.3441 | 0.038178 | 0.061335 | 0.055840 | 0.776107 |
| 4 | 34.7829 | 34.8355 | 35.1103 | 35.7047 | 0.042090 | 0.055411 | 0.054192 | 0.768127 |

Comparing other methods against each other, we found that one could perform better than expected or not depending on the image interpolated and in all the cases none of them achieved a higher PSNR value than the proposed NNV. However, the mentioned conventional algorithms are all faster than the proposed NNV. For example, in case where the interpolation ratio = 4, the NNV was about 21.2 times slower than the fastest NNI whereas in the case of the ratio = 2, the NNV became approximately 18.8 times slower than the NNI. In fact, the best interpolation method for one size of enlargement may not necessarily be the best method for a different size, in terms of the visual resolution, PSNR value and MET value. Please note that the images presented, in the experimental part of this paper, have lost some of their quality when they were reduced to fit in the paper format. This fairly reduces the differences between the presented/interpolated images. With reference to the experimental results obtained, we suggest that the proposed NNV method be recommended for further applications, especially where some image tissues, or particular details, need to be seen in their richest and most pleasant way as well as where a balmy computational cost is not an issue. Future developments of the proposed approach may be guided by techniques using higher order polynomials to interpolate.

## ACKNOWLEDGMENT

This work was supported by National Anti-counterfeit Engineering Research Center and National Natural Science Foundation of China (N0: 60772091). Rukundo Olivier and Cao Hanqiang would like to use this opportunity to thank the reviewers and editor for the helpful comments and decision, respectively.

## AUTHORS PROFILE

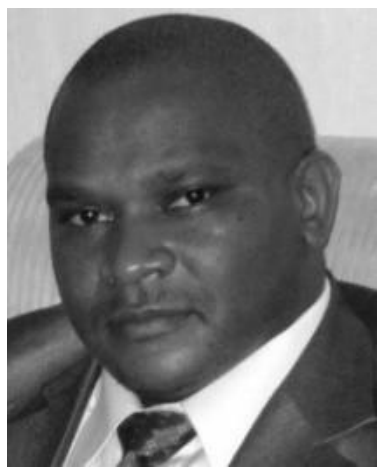

**Olivier Rukundo** born December 1981 in Rwanda currently holds B.Sc. (2005) in electronics and telecommunication engineering and M.E. (2009) in communication and information system from Kigali Institute of Science and Technology and Huazhong University of Science and Technology, respectively.
He is currently doing research in the area of signal processing at the Laboratory for Information Security and Identity, National Anti-Counterfeit Engineering Research Center and his previous published works included novel digital image interpolation algorithms as well as analog circuits test modes. Mr. Rukundo was employed for 18 months at Société Interbancaire de Monétique et de Télécompensation au Rwanda before moving on to Huazhong University of Science and Technology where he is currently a PhD candidate and expecting to graduate in June 2012.

**Dr. Hanqiang Cao** is professor currently in the department of Electronics and Information Engineering, Huazhong University of Science and Technology. His areas of expertise and interest are signal/image processing, information security.